\documentclass{revtex4-2}
\usepackage[utf8]{inputenc}
\usepackage{amsbsy}
\usepackage{amsopn}
\usepackage{amstext}
\usepackage{amsmath,amsthm,amsfonts,amssymb}
\usepackage[mathcal]{eucal}
\usepackage{mathrsfs}
\usepackage{braket}
\usepackage[english]{babel}
\usepackage{color}
\usepackage{esint}
\usepackage{graphicx}
\usepackage{float}
\usepackage{units}
\usepackage{blindtext}
\usepackage{hyperref}
\usepackage{textcomp}
\usepackage{orcidlink}

\DeclareGraphicsExtensions{.png,PNG,.pdf,.PDF}
\usepackage{hyperref}
\usepackage{slashed}
\newcommand{\ie}{\begin{equation}}
\newcommand{\fe}{\end{equation}}
\newcommand{\se}{\begin{eqnarray}}
\newcommand{\ff}{\end{eqnarray}}
\begin{document}

\title{Comment on ``Galvano-rotational effect induced by electroweak interactions in pulsars''}
\author{R. R. S. Oliveira\,\orcidlink{0000-0002-6346-0720}}
\email{rubensrso@fisica.ufc.br}
\affiliation{Departamento de F\'isica, Universidade Federal da Para\'iba, Caixa Postal 5008, 58051-900, Jo\~ao Pessoa, PB, Brazil}


\date{\today}

\begin{abstract}

In this comment, we show that the fundamental equation worked by Dvornikov in his paper is incorrect, which is given by the Dirac equation for a neutrino interacting with matter in a rotating frame (or curved space-time). In particular, Dvornikov incorrectly wrote/defined the interaction term (effective potentials or background matter) in a curved space-time. Consequently, one of his main results (if not the main one), namely the relativistic energy spectrum, is also incorrect (actually partially incorrect). So, starting from the correct Dirac equation with effective potentials in a curved space-time, we obtain the correct energy spectrum for a neutrino interacting with matter in a rotating frame. We observe that the square root of this spectrum is very different from that obtained by Dvornikov. In fact, in our case, the spectrum was obtained from a Bessel equation subject to a hard-wall confining potential, while the Dvornikov spectrum was obtained from an associated Laguerre equation. However, we note that the only correct thing in the Dvornikov spectrum are the terms outside the square root, which agree with those of the correct spectrum.

\end{abstract}

\maketitle

\section{Introduction}

In a paper published in the Journal of Cosmology and Astroparticle Physics (JCAP), entitled ``Galvano-rotational effect induced by electroweak interactions in pulsars'', Dvornikov \cite{Dvornikov} studied the electroweakly interacting particles in rotating matter, where the existence of the electric current along the axis of the matter rotation was predicted in this system. According to Dvornikov \cite{Dvornikov}, this new Galvano-rotational effect (GRE) is caused by the parity-violating interaction between massless charged particles in the rotating matter. To do such a study, Dvornikov \cite{Dvornikov} worked with the fundamental equation of relativistic quantum mechanics for spin-1/2 particles in curved space-times, that is, with the Dirac equation for a neutrino interacting with matter in a rotating frame (with constant angular velocity and in cylindrical coordinates), where the formalism used to write this equation in a curved space-time (or rotating frame) was the well-known vierbein vectors formalism (of general relativity of Einstein). Therefore, this equation includes the noninertial effects generated by a uniformly rotating frame. So, by defining a new spinor for the system, Dvornikov \cite{Dvornikov} transformed the Dirac equation into a second-order (ordinary) differential equation. Considering the limit $m\to 0$ (i.e. ultrarelativistic or massless neutrinos), Dvornikov \cite{Dvornikov} solved the differential equation and obtained the relativistic energy spectrum (or quantized energy eigenvalues) for the neutrino, where such a result was one of the most important (if not the greatest) of his paper. In fact, it is through this spectrum (with some conditions) that Dvornikov \cite{Dvornikov} obtained the induced electric current (hydrodynamic current) along the rotation axis, where this phenomenon was called by him the new galvano-rotational effect. In particular, this paper is well-written and addresses a very interesting topic involving Dirac neutrinos interacting with matter in a rotating frame. As for the formalism used, it is also very important in the literature when working with Dirac fermions in curved space-times \cite{Bakke,Bakke1,Bakke2,Bakke3,Bakke4,Cuzinatto,O1,O2,O3,O4,O5,O6,O,Andrade,Cunha,JHEP,JHEP2}.

However, analyzing in detail a paper recently published in Physical Review D (PRD) \cite{Bandyopadhyay}, we verified that the Dirac equation worked/used by Dvornikov \cite{Dvornikov} in his paper is incorrect, that is, the Dirac equation for a neutrino interacting with matter in a rotating frame (or curved space-time) was written incorrectly. To be even more specific, Dvornikov \cite{Dvornikov} incorrectly wrote/defined the interaction term (effective potentials or background matter) in a curved space-time. Consequently, one of his main results (if not the main one), namely the relativistic energy spectrum (or quantized energy eigenvalues), is also incorrect. Therefore, using Ref. \cite{Bandyopadhyay} as a basis (as well as other related papers), the present comment has as its goal to show in detail the correct Dirac equation with effective potentials in a curved space-time and, consequently, to obtain the correct energy spectrum for a neutrino interacting with matter in a rotating frame. To achieve the quantization of energy (or relativistic bound states), we will consider that our second-order differential equation is subject to a hard-wall confinement potential (analogous to the well-known finite/infinite square potential well). Without this potential, the quantization of the energy (or bound states) is not achieved. That is, unlike the Dvornikov differential equation \cite{Dvornikov}, here we will obtain a Bessel equation.


\section{Quick review of the main steps that Dvornikov took to obtain the energy spectrum}

According to Dvornikov \cite{Dvornikov}, the Dirac equation for a massive neutrino (of mass $m$) involved in the parity-violating interaction and moving in a (3+1)D curved space-time has the form \cite{Dvornikov}:
\begin{equation}\label{1}
[i\gamma^\mu (x)\nabla_\mu-m]\psi=\gamma_\mu (x)\left\{\frac{V^\mu_L}{2} [1-\gamma^5 (x)]+\frac{V^\mu_R}{2} [1+\gamma^5 (x)]\right\}\psi,
\end{equation}
where $\gamma^\mu (x)=e^{\ \mu}_a \gamma^a$ and $\gamma_\mu (x)=e^a_{\ \mu}\gamma_a$ are the curved gamma matrices, $\gamma^a$ and $\gamma_a$ are the flat gamma matrices, $e^{\ \mu}_a$ is the vierbein and $e^a_{\ \mu}$ is its inverse (and satisfy the orthogonality condition $e^a_{\ \mu} e^{\ \mu}_b =\delta^a_b$), $\nabla_\mu=\partial_\mu+\Gamma_\mu$ is the covariant derivative, $\Gamma_\mu=-\frac{i}{4}\sigma^{ab}\omega_{ab\mu}=-\frac{i}{4}\sigma^{ab}e^{\ \nu}_a e_{b\nu;\mu}$ is the spin connection, $\sigma^{ab}=\frac{i}{2}[\gamma^a,\gamma^b]$ are the generators of the Lorentz transformations in a locally Minkowskian frame, $\gamma^5 (x)=-\frac{i}{4!}E^{\mu\nu\alpha\beta}\gamma_\mu (x)\gamma_\nu (x)\gamma_\alpha (x) \gamma_\beta (x)$ is the curved fifth gamma matrix, $E^{\mu\nu\alpha\beta}=\frac{1}{\sqrt{-g}}\varepsilon^{\mu\nu\alpha\beta}$ is the covariant antisymmetric tensor in curved space-time, $g$ is the determinant of the metric (i.e. $g=$ det$(g_{\mu\nu})$), $V^\mu_{L,R}$ are the effective potentials (which depend on the electrons, neutrons or protons densities), with $V^\mu_L\neq V^\mu_R$ (since the electroweak interaction violates parity), i.e. the last term of Eq. \eqref{1} is the background matter or simply the matter term, and $\psi$ is the four-component Dirac spinor, respectively. However, unlike $\gamma^\mu (x)$, the effective potentials do not depend on the vierbein: $V^\mu_{L,R}\neq V^\mu_{L,R}(x)$ or better $V^\mu_{L,R}\neq e^{\ \mu}_a V^a_{L,R}(x)$ \cite{Dvornikov}.

According to Dvornikov \cite{Dvornikov}, the line element in the rotating frame is written as:
\begin{equation}\label{interval1}
ds^2=g_{\mu\nu}dx^\mu dx^\nu=(1-\omega^2 r^2)dt^2-dr^2-2\omega r^2 dt d\phi-r^2 d\phi^2-dz^2,
\end{equation}
where $x^\mu=(t,r,\phi,z)$ is the position four-vector in cylindrical coordinates, $\omega$ is the constant angular velocity, and $g_{\mu\nu}$ is the metric tensor, respectively. In addition, one can check that the metric tensor in Eq. \eqref{interval1} can be diagonalized (via $\eta_{ab}=e_a^{\ \mu}e_b^{\ \nu}g_{\mu\nu}$) if we use the following vierbein vectors (or vierbeins) \cite{Dvornikov}:
\begin{eqnarray}\label{vierbein1}
&& e_0^{\ \mu}=\left(1,0,-\omega,0\right),
\nonumber\\
&& e_1^{\ \mu}=\left(0,1,0,0\right),
\nonumber\\
&& e_2^{\ \mu}=\left(0,0,\frac{1}{r},0\right),
\nonumber\\
&& e_3^{\ \mu}=(0,0,0,1),
\end{eqnarray}
where $\eta_{ab}$=diag$(+1,-1,-1,-1)$ is the metric in a locally Minkowskian frame. With this, the only non-zero component of the connection one-form $\omega_{ab}=\omega_{ab\mu}dx^\mu$ is given as follows \cite{Dvornikov} (originally obtained by Ref. \cite{Bakke}):
\begin{equation}
\omega_{12\mu}=-\omega_{21\mu}=(\omega,0,1,0),
\end{equation}
where implies that $i\gamma^\mu (x)\Gamma_\mu=\frac{i}{2r}\gamma^1$ \cite{Dvornikov,Bakke}.

From this, Dvornikov \cite{Dvornikov} rewrote Eq. \eqref{1} in the following equation (with $V^\mu_{L,R}=(V^0_{L,R},0,0,0)=(V_{L,R},0,0,0)$ and $\gamma^5(x)=\gamma^5=i\gamma^0\gamma^1\gamma^2\gamma^3$, i.e. $V^\mu_{L,R}$ and $\gamma^5(x)$ do not depends on vierbein or metric \cite{Noble}):
\begin{equation}\label{2}
\mathcal{D}\psi=(\gamma^0-\omega r \gamma^2)(V_V-V_A\gamma^5)\psi,
\end{equation}
where
\begin{equation}\label{3}
\mathcal{D}=\left[i\gamma^0(\partial_0-\omega\partial_\phi)+i\gamma^1 \left(\partial_r+\frac{1}{2r}\right)+i\gamma^2\frac{\partial_\phi}{r}+i\gamma^3 \partial_z-m\right],
\end{equation}
being $V_{V,A}=(V_L \pm V_R)/2$ the vector and axial parts of the effective potentials, and the matrices $\gamma^0$, $\gamma^k$ $(k=1,2,3)$, and $\gamma^5$ are written as (chiral representation):
\begin{equation}\label{matrices}
\gamma^0=\left(
    \begin{array}{cc}
      0\ &  -1 \\
      -1\ & 0 \\
    \end{array}
  \right), \ \  \gamma^k=\left(
    \begin{array}{cc}
      0 & \sigma^k \\
      -\sigma^k & \ 0 \\
    \end{array}
  \right), \ \  \gamma^5=\left(
    \begin{array}{cc}
      1 & \ 0 \\
      0 & -1 \\
    \end{array}
  \right).
\end{equation}

So, considering the following spinor:
\begin{equation}\label{spinor}
\psi=\text{exp}(-iEt+iJ_z\phi+ip_z z)\psi_r, 
\end{equation}
Dvornikov \cite{Dvornikov} obtained:
\begin{equation}\label{4}
\left[\gamma^a Q_a-m+V\right]\psi_r=0,
\end{equation}
where $J_z=\pm\frac{1}{2},\pm\frac{3}{2},\pm\frac{5}{2},\ldots$ is the angular quantum number (i.e. a quantum number associated with the angular part of the spinor), $Q^a=q^a-q_{\text{eff}}A^{a}_{\text{eff}}$, being $q_{\text{eff}}$ the effective electric charge, $q^a=\left(E+J_z \omega-V_V,-i\partial_r,0,p_z\right)$, $V=V_A(\gamma^0-\omega r \gamma^2)\gamma^5$, and $A^{a}_{\text{eff}}$ is the potential of the effective electromagnetic field, given as follows:
\begin{equation}\label{5}
A^{a}_{\text{eff}}=\left(0,\frac{i}{2q_{\text{eff}}r},\frac{1}{q_{\text{eff}}}\left[V_V \omega r-\frac{J_z}{r}\right],0\right).
\end{equation}

Besides, Dvornikov \cite{Dvornikov} considered the solution of Eq. \eqref{4} in the form: $\psi_r=[\gamma^a Q_a+m-V]\Phi$, where $\Phi=\Phi(r)$ is a new spinor. With respect to the form of $\psi_r$, it aims to arrive at a ``quadratic Dirac equation'', that is, arriving at a second-order differential equation without passing directly by the first-order differential equations coupled with the spinor components, such as is done in Refs. \cite{Bakke,Bakke1,Bakke2,Bakke3,Bakke4,Cuzinatto,O1,O2,O3,O4,O5,O6,Cunha}. In particular, this method for obtaining a ``quadratic Dirac equation'' also has great relevance in the literature, where several papers have already used it \cite{Andrade,Gavrilov1,Gavrilov2,Vakarchuk,Peres,O7,O8,O9,O10}.

Therefore, using \eqref{5} and a the form of $\psi_r$ (defined just above), Dvornikov \cite{Dvornikov} obtained the following second-order differential equation (or ``quadratic Dirac equation'') for the spinor $\Phi$:
\begin{eqnarray}\label{6}
&& \left[\left(\partial_r+\frac{1}{2r}\right)^2+(E+J_z \omega-V_V)^2-\left(\frac{J_z}{r}-V_V \omega r\right)^2+2V_A \gamma^5\left[\left(\frac{J_z}{r}-V_V \omega r\right)\omega r-(E+J_z\omega-V_V)+\frac{\omega}{2}\Sigma_3\right]\right]\Phi,
\nonumber\\
&& +\left[\left(V_V \omega +\frac{J_z}{r^2}\right)\Sigma_3-p^2_z+V_A^2(1-\omega^2 r^2)+2m V_A (\gamma^0-\omega r \gamma^2)\gamma^5-m^2\right]\Phi=0,
\end{eqnarray}
where $\Sigma_3=\gamma^0 \gamma^3 \gamma^5$.

So, according to Dvornikov \cite{Dvornikov}, the solution of Eq. \eqref{6} can be found for ultrarelativistic particles (i.e. in the limit $m\to 0$). With this, it is possible write $\Phi$ in the form $\Phi=\upsilon\varphi$, where $\upsilon$ is a constant spinor (and satisfies $\Sigma_3\upsilon=\sigma\upsilon$ and $\gamma^5\upsilon=\chi\upsilon$, where $\sigma=\chi=\pm 1$) and $\varphi=\varphi (r)$ is a scalar function \cite{Dvornikov}. Besides, Dvornikov \cite{Dvornikov} first studied left-handed particles where $(1+\gamma^5)\psi=0$ or $V_R=0$ (with $V_{V,A}=V_L/2$), which corresponds to $\chi=+1$ (the case of right particles with $\chi=-1$ can be studied analogously). In this way, using a new variable in Eq. \eqref{6} given by $\rho=\vert V_L \vert\omega r^2$ (i.e. making a change of variable), Dvornikov \cite{Dvornikov} obtained the following equation for $\varphi_\sigma$:
\begin{equation}\label{7}
\left[\rho\partial^2_{\rho}+\partial_{\rho}-\frac{1}{4\rho}\left(l+\frac{\sigma}{2}\text{sgn}(V_L)-\frac{1}{2}\right)^2-\frac{\rho}{4}-\frac{1}{2}\left(l+\frac{\sigma}{2}\text{sgn}(V_L)-\frac{1}{2}\right)+\kappa\right]\varphi_\sigma=0,
\end{equation}
where 
\begin{equation}\label{k}
\kappa=\frac{1}{4\vert V_L\vert\omega}\left[E^2+2E(J_z \omega-V_L)-p^2_z-m^2+(V_L)^2+J^2_z\omega^2+V_L\omega \sigma\right]+\frac{1}{2}\left(l+\frac{\sigma}{2}\text{sgn}(V_L)-\frac{1}{2}\right),
\end{equation}
being $J_z=(1/2-l)\text{sgn}(V_L)$, where $l=0,\pm 1,\pm 2,\ldots$.

According to Dvornikov \cite{Dvornikov}, the solution of Eq. \eqref{7} (which vanishes at infinity, i.e. $\varphi_\sigma \to 0$ when $r\to\infty$) depends on the associated Laguerre polynomials; consequently, the parameter $\kappa$ must be equal to a positive integer, that is, $\kappa=N$, where $N=0,1,2,\ldots$ is the radial quantum number (i.e. a quantum number associated with the radial/spatial part of the spinor). Therefore, using the condition $\kappa=N$, Dvornikov \cite{Dvornikov} obtained the following relativistic energy spectrum for the Dirac neutrino (written in terms of $l$):
\begin{equation}\label{spectrum1}
E=V_L+\left(l-\frac{1}{2}\right)\omega\text{sgn}(V_L)\pm \mathcal{E}, \ \ \mathcal{E}=\sqrt{p^2_z+m^2+4\vert V_L\vert N\omega}.
\end{equation}

According to the spectrum above, we think very strange the presence of the (quadratic) mass term in the spectrum (given by $m^2$), that is, since Dvornikov \cite{Dvornikov} said that the solution of Eq. \eqref{6} can be found for ultrarelativistic or massless particles (i.e. $m\to 0$), such a term was not supposed to be present in the spectrum. Besides, it is important to mention that the presence of sgn$(V_L)$ in $J_z$ is also something very strange, that is, the correct according to Ref. [16] of Dvornikov's paper \cite{Dvornikov} should be only $J_z=l+1/2$ (in fact, this is used in Ref. [15] as $j=l+1/2$, which also cites [16]. Therefore, Dvornikov \cite{Dvornikov} should have cited [15,16] when introduced $J_z$ ($j$ would be better) in his paper).


\section{The correct Dirac equation with effective potentials in a curved space-time for a neutrino interacting with matter in a rotating frame}

Here, let us get the correct Dirac equation with effective potentials in a curved space-time for a neutrino interacting with matter in a rotating frame. So, according to Ref. \cite{Bandyopadhyay}, the Dirac action for a neutrino interacting with matter in a curved space-time is given by
\begin{equation}\label{S}
S_{int}=\int d^4x \sqrt{-g}\mathcal{L}_{int},
\end{equation}
where $\mathcal{L}_{int}$ is the interaction Lagrangian density (due to the charged current interaction effects), given as follows
\begin{equation}\label{L}
\mathcal{L}_{int}=-\frac{G_F}{\sqrt{2}}J_\mu\Bar{\nu}_e \gamma^\mu (1-\gamma^5)\nu_e, \ \ (\mu=t,r,\Tilde{\theta},\varphi),
\end{equation}
being $G_F$ the Fermi coupling constant, $J_\mu=\Bar{\psi}_e \gamma_\mu (1-\gamma^5)\psi_e$ is the charged current due to the electron field $\psi_e$ (i.e. Dirac spinor of the electron), $\gamma_\mu=\gamma_\mu(x)=e_{\ \mu}^b\gamma_b$ or $\gamma^\mu=\gamma^\mu(x)=e^\mu_{\ a}\gamma^a$ are curved gamma matrices, and $\nu_e$ is the electron neutrino field (i.e. Dirac spinor of the electron neutrino). Indeed, using the standard prescription (``standard recipe') widely used in the literature to transform the flat Dirac equation into the curved Dirac equation, i.g. $d^4x \to d^4x\sqrt{-g}$, $\gamma^\mu\to \gamma^\mu=e^\mu_{\ a}\gamma^a$ and $\partial_\mu \to \nabla_\mu=\partial_\mu+\Gamma_\mu$ \cite{Bakke,Bakke1,Bakke2,Bakke3,Bakke4,Cuzinatto,O1,O2,O3,O4,O5,O6,O,Andrade,Cunha,JHEP,JHEP2,Noble,Lawrie}, into flat interaction Lagrangian given in Refs. \cite{K1,K2,Akhmedov,Pal} (with $\mathcal{H}_{int}=-\mathcal{L}_{int}$), we get exactly \eqref{L} (or \eqref{S}). Therefore, the curved interaction Lagrangian \eqref{L} (or curved action \eqref{S}) is indeed correct.

So, according to Ref. \cite{Bandyopadhyay}, in a static and spherically symmetric space-time (but can be cylindrical too), we can approximate the background electron current as $J_\mu=n_e e^0_{\ \mu}$, where $n_e=n_e (x)$ is the electron number density or simply the electron density. Consequently, the interaction Lagrangian \eqref{L} reduces to \cite{Bandyopadhyay} (we believe this reference forgot the minus sign)
\begin{equation}\label{L1}
\mathcal{L}_{int}=-\frac{G_F}{\sqrt{2}}J_t\Bar{\nu}_e \gamma^t (1-\gamma^5)\nu_e=-\frac{G_F}{\sqrt{2}}n_e e^0_{\ t}\Bar{\nu}_e e^t_{\ 0}\gamma^0 (1-\gamma^5)\nu_e=-\frac{G_F}{\sqrt{2}}n_e\Bar{\nu}_e\gamma^0 (1-\gamma^5)\nu_e=-(\sqrt{2}G_F n_e)[\Bar{\nu}_e\gamma^0 P_L\nu_e],
\end{equation}
where $P_L=\frac{1}{2}(1-\gamma^5)$ is the left-handed projection operator, and we use the fact that $e^0_{\ t}e^t_{\ 0}=\delta^t_t=\delta^0_0=1$. Therefore, we see that $\mathcal{L}_{int}$ does not contain any explicit dependence on vierbein or space-time metric (i.e. $\mathcal{L}_{int}^{curved}=\mathcal{L}_{int}^{flat}$) \cite{Bandyopadhyay} (however, the metric dependence is contained in the term $\sqrt{-g}$ in \eqref{S} \cite{Bandyopadhyay}).

Now, according to Ref. \cite{Bandyopadhyay}, the dynamics of a free fermion in a curved space-time is governed by the following action
\begin{equation}\label{S2}
S_{\psi}=\int d^4x \sqrt{-g}\mathcal{L}_{\psi},
\end{equation}
where $\mathcal{L}_{\psi}$ is the free Dirac Lagrangian density (or free Lagrangian density), given as follows
\begin{equation}\label{L3}
\mathcal{L}_{\psi}=-\Bar{\psi}[i\gamma^\mu D_\mu+m]\psi, 
\end{equation}
being $D_\mu=\nabla_\mu=\partial_\mu+\Gamma_\mu$ the (spin) covariant derivative, $\Gamma_\mu$ is the spin connection, and $m$ is the mass term (of the fermion). However, according to Ref. \cite{Lawrie} (page 212), the minus sign in \eqref{L3} is in the wrong place, i.e. the minus sign should be in the mass term, such as $\mathcal{L}_{\psi}=\Bar{\psi}[i\gamma^\mu D_\mu-m]\psi$ (we believe this is the correct Lagrangian). So, doing $\Bar{\nu}_e\to \Bar{\psi}$ and $\nu_e \to \psi$ (i.e. consider only a single neutrino flavor where the mixing angle is null/zero), implies that the total/complete Lagrangian for the neutrino (electron neutrino, actually) will be given by (with the negative signs already corrected)
\begin{equation}\label{L4}
\mathcal{L}_{total}=\Bar{\psi}\left[i\gamma^\mu D_\mu-m-(\sqrt{2}G_F n_e)\gamma^0 P_L\right]\psi. 
\end{equation}

Therefore, using the following Euler-Lagrange equation \cite{Greiner}
\begin{equation}
\partial_\mu\frac{\partial\mathcal{L}_{total}}{\partial(\partial_\mu \Bar{\psi})}-\frac{\partial\mathcal{L}_{total}}{\partial_\mu \Bar{\psi}}=0,
\end{equation}
we can obtain the following Dirac equation for a (left-handed) neutrino interacting with matter in a curved space-time
\begin{equation}\label{L5}
\left[i\gamma^\mu D_\mu-m-(\sqrt{2}G_F n_e)\gamma^0 P_L\right]\psi=0, 
\end{equation}
or yet
\begin{equation}\label{L6}
\left[i\gamma^\mu (x) D_\mu-m\right]\psi=\gamma_\mu(x) V^\mu_L(x) P_L\psi, 
\end{equation}
where we define $V^\mu_L(x)=e^{\ \mu}_a V^a_L$ (left-handed effective potential), with $V^L_a=(V^0_L,0,0,0)=(\sqrt{2}G_F n_e,0,0,0)$, and $\gamma_\mu=e^a_{\ \mu}\gamma_a$ (in particular, a more general case where $n_e\to n_e -n_n/2$, being $n_n$ the neutrons density, is give by \cite{Akhmedov,Pal}). However, to also include a right-handed neutrino in the equation, simply do: $V^\mu_L(x) P_L (x)\to V^\mu_L P_L(x)+V^\mu_R(x) P_R$, where $V^\mu_R(x)=e^{\ \mu}_a V^a_R$ (with $V^a_R=(V^0_R,0,0,0)$) would be the right-handed effective potential (with $V^\mu_R(x)\neq V^\mu_L(x)$ or $V^0_R\neq V^0_L$), being $P_R=\frac{1}{2}(1+\gamma^5)$ the right-handed protection operator (in particular, considering that $n_e$ and $n_n$ do not depend on the coordinates, i.e. are uniform quantities, implies that theses effective potentials are the same worked by Dvornikov \cite{Dvornikov} in his Table 1). Therefore, the correct Dirac equation for Dvornikov's paper \cite{Dvornikov} (Dirac equation with effective potentials in a curved space-time) should be given by
\begin{equation}\label{L7}
[i\gamma^\mu (x)\nabla_\mu (x)-m]\psi=\gamma_\mu (x)\left\{\frac{V^\mu_L(x)}{2} [1-\gamma^5 (x)]+\frac{V^\mu_R (x)}{2} [1+\gamma^5 (x)]\right\}\psi,
\end{equation}
where both $\gamma_\mu (x)V^\mu_L(x)$ and $\gamma_\mu (x)V^\mu_R(x)$ (and $\gamma^5 (x)$) do not depend on vierbein or metric ($\gamma_\mu (x)V^\mu_{L/R}(x)=\gamma_0 V^0_{L/R}$). Before proceeding, it is important to mention that although Dvornikov also worked with Eq. \eqref{1} (incorrect equation) in another paper (given by Ref. \cite{JHEP}), the result still remained correct (but incomplete, as shown by Ref. \cite{JHEP2}) since the slow rotation regime was considered (which implies that $e_{\ t}^0\approx 1$). In this case (slow rotation), it makes no difference whether $g^\mu$ depends on $e^\mu_a$ or not (again, the correct thing is to depend on $e^\mu_a$). However, we will see that here it makes a significant difference, that is, unlike \cite{JHEP}, here, Dvornikov \cite{Dvornikov} chose another way of including rotation in the system. Consequently, we will see that the Dirac equation \eqref{L7} will generate another type of second-order differential equation (which is different from Refs. \cite{JHEP,Dvornikov}, where an associated Laguerre equation was generated).

So, as a consequence of Eq. \eqref{L7}, Eq. \eqref{2} should be written as
\begin{equation}\label{L8}
\mathcal{D}\psi=\gamma^0(V_V-V_A\gamma^5)\psi,
\end{equation}
whereas \eqref{4} should be
\begin{equation}\label{L9}
\left[\gamma^a \tilde{Q}_a-m+\tilde{V}\right]\psi_r=0,
\end{equation}
where $\tilde{Q}^a=q^a-q_{\text{eff}}\tilde{A}^{a}_{\text{eff}}$, being $q^a=\left(E+J_z \omega-V_V,-i\partial_r,0,p_z\right)$, $\tilde{V}=V_A\gamma^0\gamma^5$, and $\tilde{A}^{a}_\text{eff}=\left(0,\frac{i}{2q_{\text{eff}}r},-\frac{1}{q_{\text{eff}}}\frac{J_z}{r},0\right)$.

Now, defining $\psi_r\equiv [\gamma^a\tilde{Q}_a+m-\tilde{V}]\Phi$, we can obtain from Eq. \eqref{L9} a second-order differential equation for the spinor $\Phi$, given as follows
\begin{equation}\label{L10}
\left[\left(\partial_r+\frac{1}{2r}\right)^2+(E+J_z\omega-V_V)^2-\frac{J_z^2}{r^2}-2V_A\gamma^5(E+J_z\omega-V_V)+\frac{J_z}{r^2}\Sigma_3-p^2_z+V_A^2+2m V_A\gamma^0\gamma^5-m^2\right]\Phi=0,
\end{equation}
or better
\begin{equation}\label{L10}
\left[\partial_r^2+\frac{1}{r}\partial_r-\frac{(J_z-\frac{\Sigma_3}{2})^2}{r^2}+(E+J_z\omega-V_V)^2-2V_A\gamma^5(E+J_z\omega-V_V)-p^2_z+V_A^2+2m V_A\gamma^0\gamma^5-m^2\right]\Phi=0,
\end{equation}
where $\Sigma_3=\gamma^0 \gamma^3 \gamma^5$ and $\Sigma_3^2=1$.

Here, such as stated/considered by Dvornikov \cite{Dvornikov}, the solution of Eq. \eqref{L10} can be found for ultrarelativistic particles ($m\to 0$); consequently, it implies that all terms of the equation containing $m$ must vanish (or be zero). If not (i.e. $m\neq 0$), then to solve the equation, it must first be diagonalized (which would be something very complex) \cite{Ferreira}. In other words, the product $\gamma^0\gamma^5$ is anti-diagonal (this can be easily shown using \eqref{matrices}), i.e. $\gamma^0\gamma^5$=anti-diag$(+1,-1)$. Therefore, here, we consider the limit $m\to 0$. Besides, we also consider left-handed neutrinos where $(1+\gamma^5)\psi=0$ or $V_R=0$, which implies $V_{V,A}=V_L/2$. In this way, using these considerations and $\Phi$ written as $\Phi=\upsilon\varphi$ (with $\Sigma_3\upsilon=\sigma\upsilon$ and $\gamma^5\upsilon=+\upsilon$, and $\sigma=\pm 1$), Eq. \eqref{L10} becomes
\begin{equation}\label{L11}
\left[\partial_r^2+\frac{1}{r}\partial_r-\frac{(J_z-\frac{\sigma}{2})^2}{r^2}+\left(E+J_z\omega-\frac{V_L}{2}\right)^2-V_L\left(E+J_z\omega-\frac{V_L}{2}\right)+\frac{V^2_L}{4}-p^2_z\right]\varphi_\sigma=0,
\end{equation}
or better
\begin{equation}\label{L12}
\left[\frac{d^2}{dr}+\frac{1}{r}\frac{d}{dr}-\frac{A_\sigma^2}{r^2}+B^2\right]\varphi_\sigma (r)=0,
\end{equation}
where we define
\begin{equation}\label{L13}
A_\sigma\equiv\Big\vert J_z-\frac{\sigma}{2}\Big\vert\geq 0, \ \ B\equiv\sqrt{(E+J_z\omega-V_L)^2-p^2_z}>0.
\end{equation}


\section{The correct energy spectrum for a neutrino interacting with matter in a rotating frame}

According to Refs. \cite{B1,B2,B3,V1,V2,ARXIV}, Eq. \eqref{L12} is the well-known Bessel equation (a type of second-order differential equation), whose solution (``radial wave function'') is given by
\begin{equation}\label{L14}
\varphi_\sigma (r)= c_1J_{A_\sigma}
(Br)+c_2Y_{A_\sigma}(Br),
\end{equation}
where $J_{A_s}(Br)$ and $Y_{A_s}(Br)$ are the Bessel functions of first and second kinds, and $c_1$ and $c_2$ are arbitrary constants. So, as at the origin $(r = 0)$ the Bessel function of first kind is finite/regular while the Bessel function of second kind is infinite/irregular, implies that: $c_1 \neq 0$ and $c_2=0$ \cite{B1,B2,B3,V1,V2,ARXIV}. Therefore, the solution \eqref{L14} becomes
\begin{equation}\label{L15}
\varphi_\sigma (r)= c_1J_{A_\sigma}
(Br).
\end{equation}
where must satisfy $\varphi_\sigma (r\to 0)= 0$, i.e., this is our first boundary condition (null function at the origin).

Now, we would like to restrict the motion of the neutrino in a region where a hard-wall confining potential is present, i.e., a finite distance where the function is also null. So, according to Refs. \cite{B1,B2,B3,V1,V2,ARXIV}, the following boundary condition (our second boundary condition) describes this kind of confinement
 \begin{equation}\label{L16}
\varphi_\sigma (r=r_0)=0,
\end{equation}
which means that the function vanishes at a fixed radius $r_0$, with $0<r_0<\infty$. According to Refs. \cite{B1,B2,B3,V1,V2,ARXIV}, a plausible physical motivation/argument for using this second boundary condition (or hard-wall confining potential) is that it allows obtaining from a Bessel equation the quantization of the energy (i.e. to obtain the bound states). In particular, we can say that this boundary condition is similar to the finite/infinite square potential well \cite{Griffiths}, which confines a non-relativistic particle in a finite region in space and, consequently, generates the energy's quantization (i.e. the confinement leads to quantization). As we will see later, our spectrum will depend on $1/r^2_0$ just as it happens in the non-relativistic spectrum for a particle subject to the finite/infinite square potential well (in fact, since here we will have a relativistic spectrum, $1/r^2_0$ will be within the square root).

Besides, assuming that $B r_0\gg 1$, the Bessel function \eqref{L15} can be written as follows \cite{B1,B2,B3,V1,V2,ARXIV}
\begin{equation}\label{L17}
J_{A_\sigma}(B r_0)\to\sqrt{\frac{2}{\pi B r_0}}\cos\left(Br_0-\frac{A_\sigma\pi}{2}-\frac{\pi}{4} \right).
\end{equation}

Consequently, the boundary condition \eqref{L16} implies that the argument of \eqref{L17} must satisfy the following relation \cite{B1,B2,B3,V1,V2,ARXIV}
\begin{equation}\label{L18}
\left(Br_0-\frac{A_\sigma\pi}{2}-\frac{\pi}{4}\right)=(2N+1)\frac{\pi}{2}, \ \ (N=0,1,2,\ldots),
\end{equation}
where $n$ is a quantum number, sometimes called the radial quantum number (since it arises from a radial differential equation). Therefore, using the quantization condition \eqref{L18} with \eqref{L13}, we obtain the following energy spectrum for a neutrino interacting with matter in a rotating frame (and subject to a hard-wall confining
potential)
\begin{equation}\label{Energy}
E=V_L+\left(l-\frac{1}{2}\right)\omega\text{sgn}(V_L)\pm \mathcal{\tilde{E}}, \ \ \mathcal{\tilde{E}}=\sqrt{p^2_z+\frac{\pi^2}{r^2_0}\left(N+\frac{\vert J_z-m_s\vert}{2}+\frac{3}{4}\right)^2}.
\end{equation}

So, comparing the correct spectrum, given by \eqref{Energy}, with the incorrect spectrum, given by \eqref{spectrum1}, we see that the main difference between them is with respect to the square root, that is, $\mathcal{\tilde{E}}\neq\mathcal{E}$. In other words, the correct thing is that the square root does not depend on the angular velocity $\omega$ (or background matter $V_L$). Thus, unlike \eqref{spectrum1}, here, in addition to the spectrum (square root) still remains quantized (due to $N$ and $J_z$) even in the absence of angular velocity ($\omega=0$) or background matter ($V_L=0$), it also depends on the angular quantum number $J_z$ and the spin magnetic quantum number $m_s=\sigma/2=\pm 1/2=\uparrow\downarrow$ \cite{ARXIV}. From the above, the only correct thing in the Dvornikov spectrum are the terms outside the square root, which agree with those of the correct spectrum.


\section{Final remarks}

In the present comment, we show that the fundamental equation worked by Dvornikov \cite{Dvornikov} in his paper is incorrect, which is given by the Dirac equation for a neutrino interacting with matter in a rotating frame (or curved space-time). In particular, Dvornikov \cite{Dvornikov} incorrectly wrote/defined the interaction term (effective potentials or background matter) in a curved space-time. That is, such potentials should also depend on the vierbein; however, this was not what happened. Consequently, one of his main results (if not the main one), namely the relativistic energy spectrum (or quantized energy eigenvalues), is also incorrect (actually partially incorrect). So, starting from the correct Dirac equation with effective potentials in a curved space-time, we obtain the correct energy spectrum for a neutrino interacting with matter in a rotating frame. We observe that the square root of this spectrum is very different from that obtained by Dvornikov \cite{Dvornikov}. In fact, in our case, the spectrum was obtained from a Bessel equation subject to a hard-wall confining potential (analogous to the well-known finite/infinite square potential well), while the Dvornikov spectrum \cite{Dvornikov} was obtained from an associated Laguerre equation. However, we note that the only correct thing in the Dvornikov spectrum \cite{Dvornikov} are the terms outside the square root, which agree with those of the correct spectrum.

\section*{Acknowledgments}

\hspace{0.5cm}
The author would like to thank the Conselho Nacional de Desenvolvimento Cient\'{\i}fico e Tecnol\'{o}gico (CNPq) for financial support through the postdoc grant No. 175392/2023-4, and also to the Department of Physics at the Universidade Federal da Para\'{i}ba (UFPB) for hospitality and support.

\section*{Data availability statement}

\hspace{0.5cm} This manuscript has no associated data or the data will not be deposited. [Author’ comment: There is no data associated with this manuscript or no data has been used to prepare it.]

\end{document}